\documentclass[11pt,dvips]{article}

\usepackage{epsfig,times} 
\usepackage{picinpar}
\usepackage{wrapfig}
\usepackage{floatflt}

\makeatletter
\renewcommand{\section}{\@startsection{section}{1}{0in}
	{0.4\baselineskip}{0.1\baselineskip}{\Large\bf}}
\renewcommand{\subsection}{\@startsection{subsection}{2}{0in}
	{0.25\baselineskip}{-\baselineskip}{\large\bf}}
\renewcommand{\subsubsection}{\@startsection{subsubsection}{3}{0in}
	{0.1\baselineskip}{-\baselineskip}{\normalsize\bf}}
\makeatother
\newcommand{\icrc}{$26^{\rm th}$ ICRC\ (Salt Lake City, 1999)}
\def\gray{$\gamma$-ray\ }
\def\grays{$\gamma$-rays\ }

\newcommand{\aap}{A\&A\ }
\newcommand{\apj}{ApJ\ }

\def\fwb{7.8cm}
\def\fwc{8.3cm}
\begin{document}

%
\makeatletter\newcommand{\ps@icrc}{
\renewcommand{\@oddhead}{Proc.~\icrc, \slshape{OG 2.4.03}\hfil}}
\makeatother\thispagestyle{icrc}
%
%

\begin{center}
{\LARGE \bf 
The Galactic contribution to high latitude diffuse \gray emission}
\end{center}

\begin{center}
{\bf Andrew W.~Strong$^1$, Igor V.~Moskalenko$^{1,2}$, and Olaf Reimer$^1$}\\
{\it $^{1}$MPI f\"ur extraterrestrische Physik, D--85740 Garching, Germany\\
$^{2}$Institute for Nuclear Physics, Moscow State University, 119 899 Moscow, 
   Russia}
\end{center}

\begin{center}
{\large \bf Abstract\\}
\end{center}
\vspace{-0.5ex}
Recent evidence for a large Galactic halo, based on cosmic-ray
radioactive nuclei, implies a significant contribution from inverse
Compton emission at high Galactic latitudes. We present predictions for
the expected intensity distribution, and show that the EGRET gamma-ray
latitude distribution is well reproduced from the plane to the poles.
We show that the Galactic component at high latitudes may be comparable
to the extragalactic emission in some energy ranges.

\vspace{1ex}

%
%
\section{Introduction:} \label{intro}
The origin of the truly extragalactic \gray background is still
unknown.  The models discussed range from the primordial black hole
evaporation (Page \& Hawking 1976) and annihilation of exotic particles
in the early Universe (Cline \& Gao 1992) to the contribution of
unresolved discrete sources such as active galaxies (Sreekumar et al.\
1998), while the spectrum of the extragalactic emission itself is
uncertain.  The latter can be addressed only by the accurate study of
the Galactic diffuse emission at high Galactic latitudes.  Moreover
there is growing evidence for a large halo contribution to the \gray
background.  An indication for a \gray halo was also found by Dixon
et al.\ (1998) from analysis of EGRET data.  Studies of $^{10}$Be
(Strong \& Moskalenko 1998) gave the range $z_h = 4-12$ kpc for
nucleons,  Webber \& Soutoul (1998) find $z_h = 2-4$ kpc from $^{10}$Be
and $^{26}$Al data,  Ptuskin \& Soutoul (1998) find $z_h =
4.9^{+4}_{-2}$ kpc.

Gamma rays provide a tracer of the electron halo via inverse Compton
(IC) emission. A study of the Galactic emission requires a systematic
approach:  computation of a realistic interstellar radiation field and
self-consistent calculation of the electron spectrum in 3D.  We use
our GALPROP model\footnote{ For interested users our model and data
sets are available in the public domain on the World Wide Web, {\tt
http://www.gamma.mpe-garching.mpg.de/$\sim$aws/aws.html} }, which has
been shown to be consistent with many kinds of data related to cosmic
ray origin and propagation, to calculate the Galactic contribution to
the high latitude diffuse \gray emission (Strong, Moskalenko, \& Reimer
1999).  The models have been described in full detail in Strong
\& Moskalenko (1998), for a recent review of our results see Strong \&
Moskalenko (1999).

\section{Description of the models:} \label{desc}
The models are three dimensional with cylindrical symmetry in the
Galaxy. For a given halo size the diffusion coefficient (as a function
of momentum) and the reacceleration parameters are determined by the
Boron-to-Carbon ratio; the momentum-space diffusion coefficient is
related to the spatial coefficient (Berezinskii et al.\ 1990). The
injection spectrum of particles is assumed to be a power law in
momentum, if necessary with a break.  The magnetic field is adjusted to
match the 408 MHz synchrotron longitude and latitude distributions. The
interstellar hydrogen and He distribution uses HI and CO surveys and
information on the ionized component.  Energy losses for particles by
ionization, Coulomb interactions, bremsstrahlung, inverse Compton, and
synchrotron are included.  The distribution of cosmic-ray sources is
adjusted (Strong \& Moskalenko 1998) to match the cosmic-ray
distribution obtained from EGRET \gray data (Strong \& Mattox 1996).
The interstellar radiation field is based on stellar population models
and COBE results, plus the cosmic microwave background (Strong,
Moskalenko, \& Reimer 1999). Inverse Compton scattering is treated as
in Moskalenko \& Strong (1999) including the effect of the anisotropy
of the ISRF, and gas related \gray intensities ($\pi^0$-decay and
bremsstrahlung) are computed using the column densities of HI and H$_2$
based on 21-cm and CO surveys.

\begin{figure}[t]
\centerline{
   \psfig{file=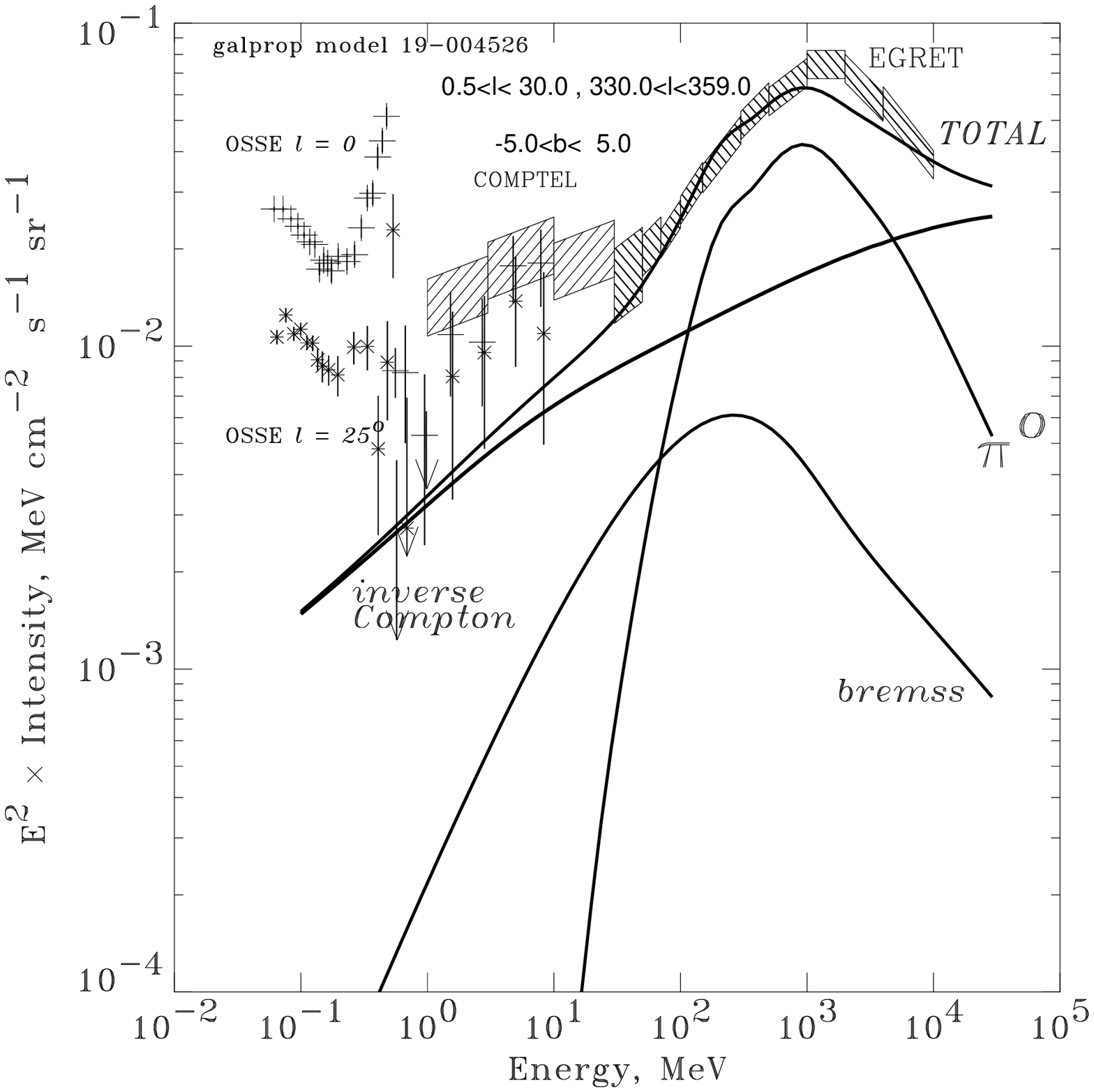,width=\fwc,clip=}\hspace{6mm}
   \psfig{file=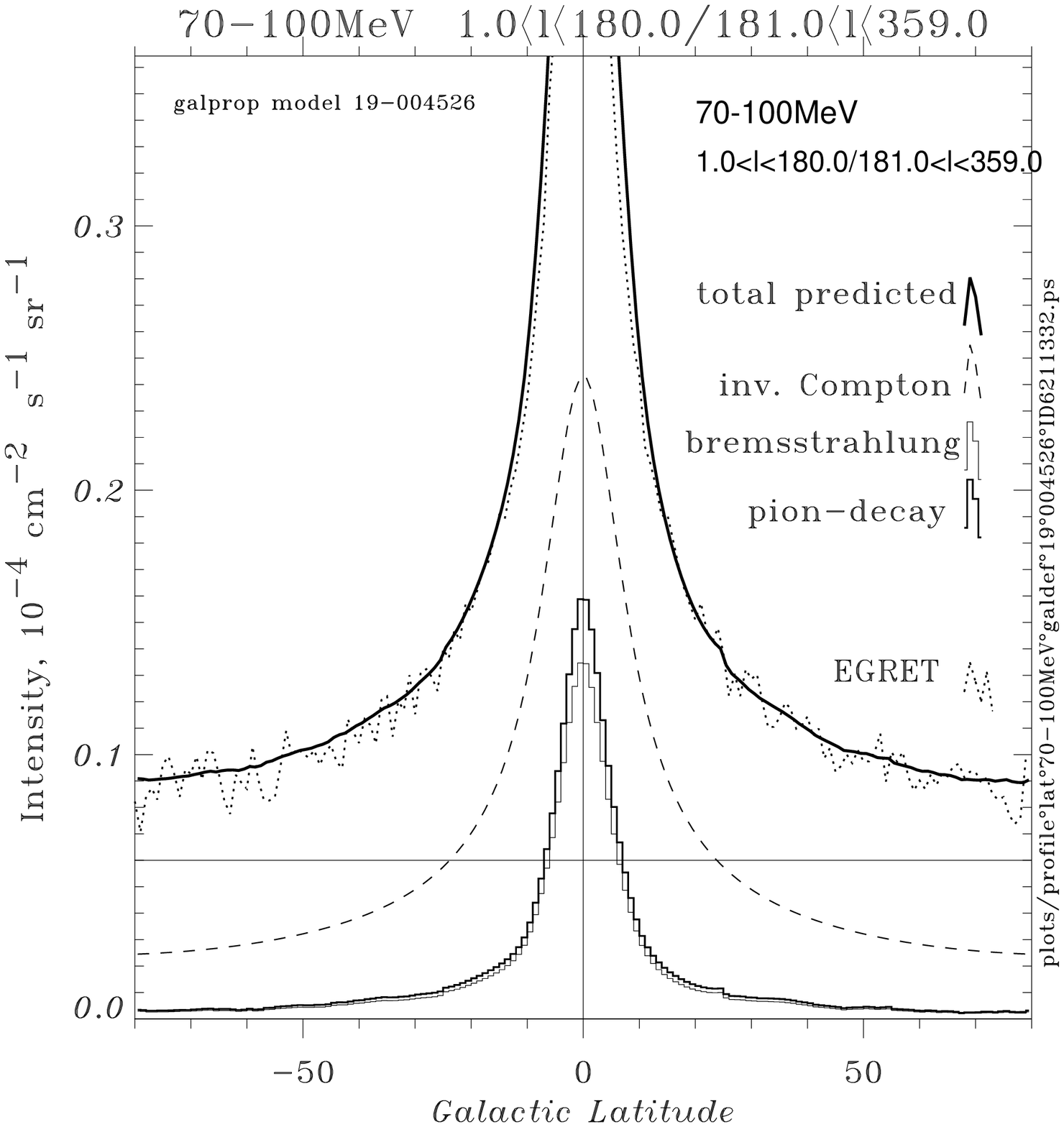,width=\fwb,clip=}\hspace{1mm}
}
\parbox{81mm}{%
\caption[OG_2.4.03_f1.ps]{ 
Gamma-ray energy spectrum of the inner Galaxy ($300^\circ \le l\le
30^\circ$, $|b|\le 5^\circ$) compared with our model calculations (electron
injection index $-1.8$, and modified nucleon spectrum).  Curves show
the contribution of IC, bremsstrahlung, and $\pi^0$-decay, and the
total.  Data: EGRET (Strong \& Mattox 1996), COMPTEL
(Strong et al.\ 1999), OSSE ($l=0, 25^\circ$: Kinzer et al.\ 1999).
\label{fig1}}
}\hspace{6mm}
\parbox{81mm}{%
\caption[OG_2.4.03_f2.ps]{
High latitude distribution (enlarged) of 70--100 MeV \grays from the
EGRET compared to our model calculation.  Separate components:
IC (dashes), bremsstrahlung (thin histogram),
$\pi^0$-decay (thick histogram), horizonal line: isotropic background.
EGRET data (point sources removed): dotted line.
\vspace{1\baselineskip}\vspace{2pt}
\label{fig2}}
}
\end{figure}

\section{Inner Galaxy:} \label{inner}
Recent results from both COMPTEL and EGRET indicate that IC scattering
is a more important contributor to the diffuse emission that previously
believed.  The puzzling excess in the EGRET data $>1$ GeV relative to
that expected for $\pi^0$-decay has been suggested to originate in IC
scattering from a hard interstellar electron spectrum (e.g., Pohl \&
Esposito 1998).  Our combined approach allows us to test this
hypothesis (Strong, Moskalenko, \& Reimer 1999).

Our first concern was to reproduce the \gray spectrum of the inner
Galaxy.  This is possible by invoking a hard electron spectrum
with injection spectral index --1.8, which after
propagation (with reacceleration) provides consistency with radio
synchrotron data.  Following Pohl \& Esposito (1998), for this model we
do {\it not} require consistency with the locally measured electron
spectrum above 10 GeV because the rapid energy losses cause a clumpy
distribution so that this is not necessarily representative of the
interstellar average.  For this case, the interstellar electron
spectrum deviates strongly from that locally measured.

Further improvement can be obtained by allowing some freedom in the
nucleon spectrum at low energies.  Some freedom is allowed since solar
modulation affects direct measurements of nucleons below 20 GeV, and
the locally measured nucleon spectrum may not necessarily be
representative of the average on Galactic scales either in spectrum or
intensity due to details of Galactic structure (e.g.\ spiral arms). By
introducing some flattening of the nucleon spectrum below 20
GeV, a small steepening above 20 GeV, and a suitable normalization, an
improved match to the inner Galaxy EGRET spectrum is indeed possible
(Fig.~\ref{fig1}).

The modified nucleon spectrum must be checked against the stringent
constraints on the {\it interstellar} spectrum provided by antiprotons
and positrons. (Such tests sample the Galactic-scale properties of CR
$p$ and He rather than just the local region, independent of
fluctuations due to local primary CR sources.) As expected the
$\bar{p}$ and $e^+$ predictions are higher than for the conventional
model (with nucleon spectrum matching the local measurements) but still
within the observational limits (Strong, Moskalenko, \& Reimer 1999).

This is our best model so far. Further tests against the \gray
longitude and latitude profiles at all the EGRET energies show a good
overall agreement (an example is shown in Fig.~\ref{fig2}).

In order to reproduce the low-energy ($<$ 30 MeV) \gray emission via
diffuse processes it is necessary to invoke an upturn of the electron
spectrum below about 200 MeV to compensate the increasing ionization
losses.  (A steep slope continuing to higher energies would violate the
synchrotron constraints on the spectral index.) However, the adoption of
such a steep low-energy electron spectrum has problems associated with
the very large power input to the interstellar medium (Skibo et
al.\ 1997), and is {\it ad hoc} with no independent supporting
evidence. Moreover the OSSE-GINGA \gray spectrum is steeper than
$E^{-2}$ below 500 keV (Kinzer et al.\ 1999) which would require an
even steeper electron injection spectrum than adopted here.  It is more
natural to consider that the COMPTEL excess is just a continuation of
the same component producing the OSSE-GINGA spectrum.  Most probably
therefore the excess emission at low energies is produced by a
population of sources such as supernova remnants,  as has been
proposed  for the diffuse hard X-ray emission from the plane observed
by RXTE (Valinia et al.\ 1998), or X-ray transients in their low state as
suggested for the OSSE diffuse hard X-rays (Lebrun et al.\ 1999).

\begin{figure}[t]
\centerline{
   \psfig{file=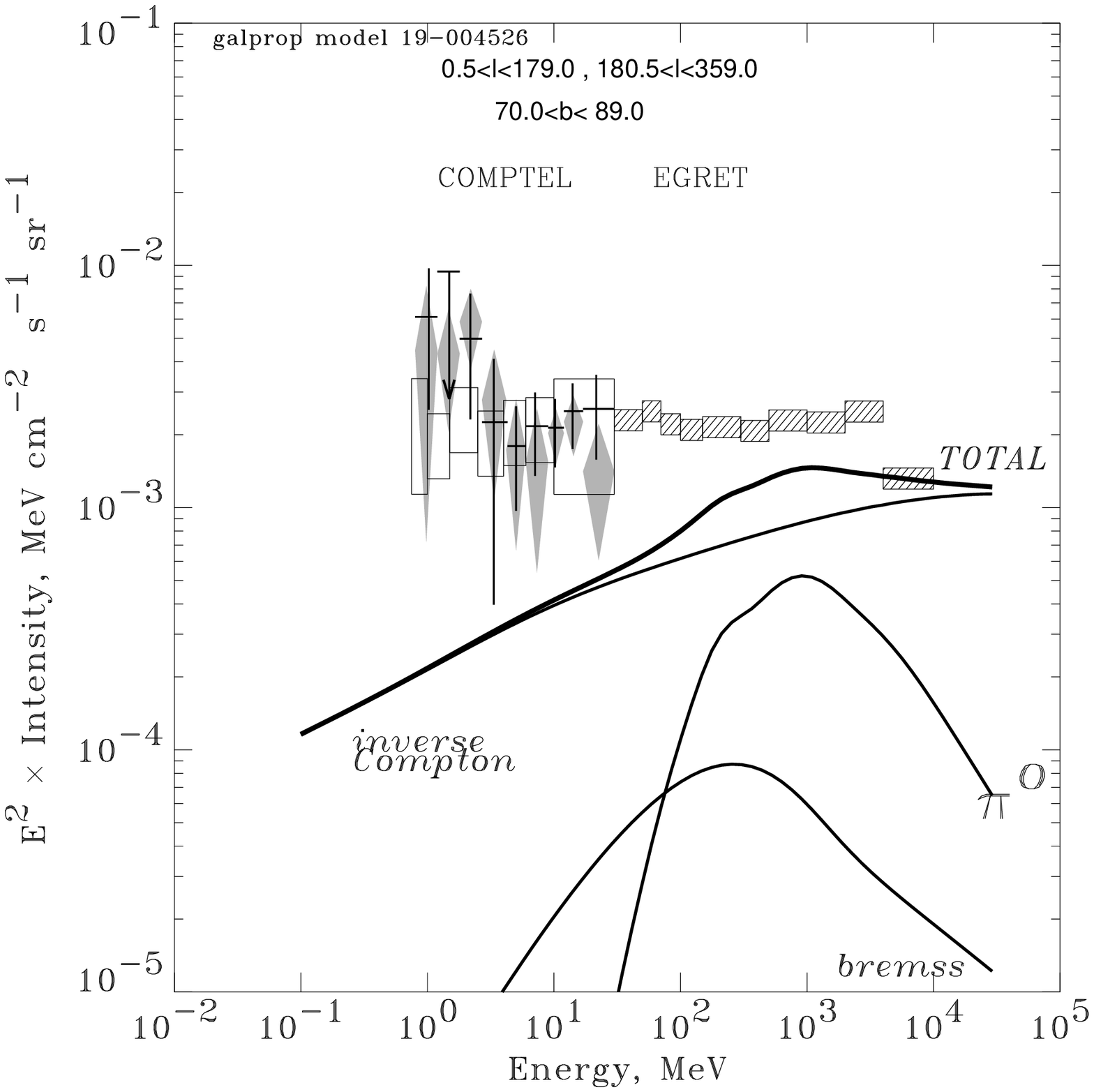,width=\fwc,clip=}
   \psfig{file=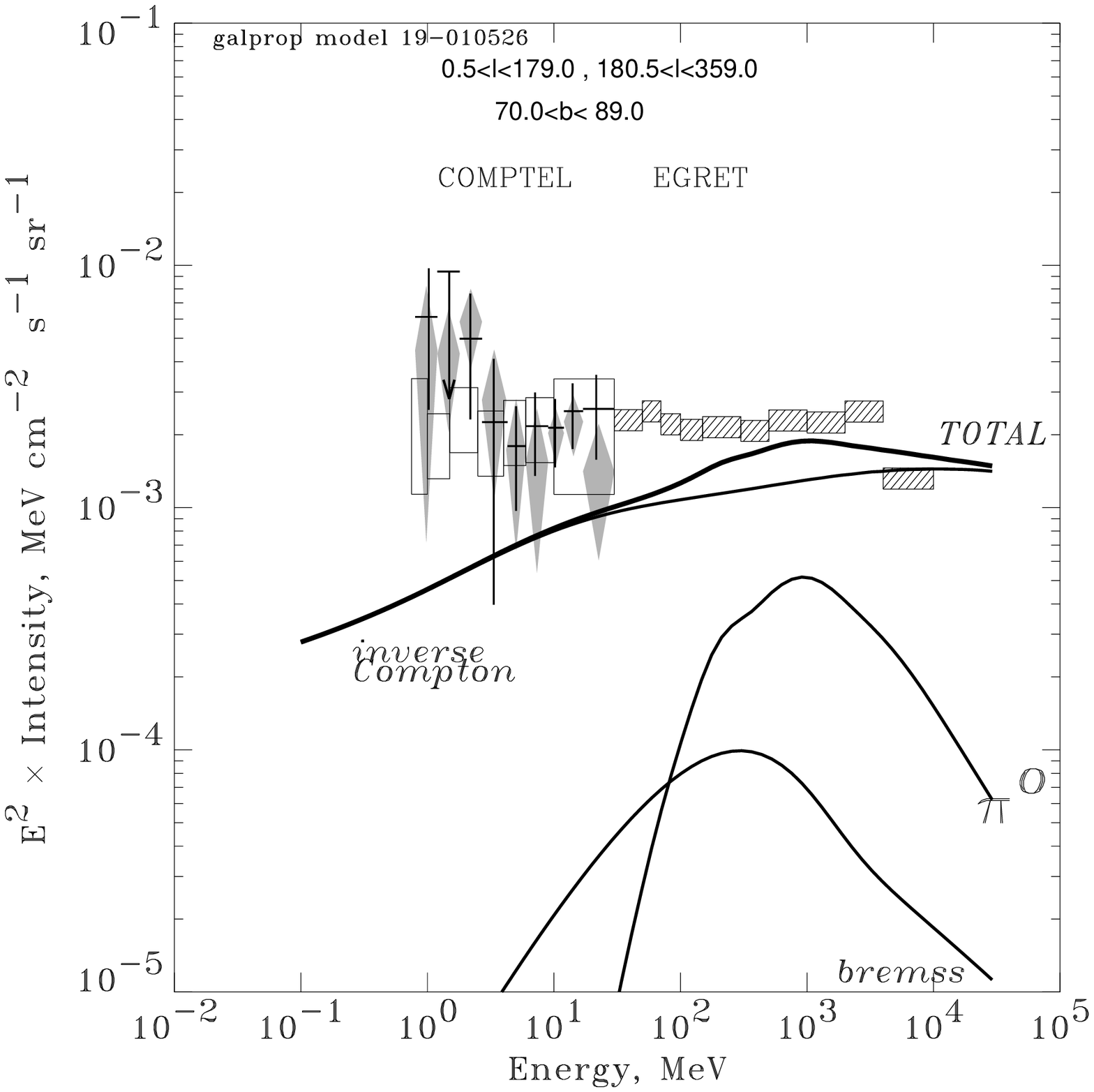,width=\fwc,clip=}
}
\caption[OG_2.4.03_f3a.ps,OG_2.4.03_f3b.ps]{
Energy spectrum of \grays from high Galactic latitudes ($|b|\ge
70^\circ$, all longitudes) for $z_h=4$ kpc (left) and $z_h=10$ kpc
(right).  Shaded areas: EGRET total intensity from Cycle 1--4 data.
COMPTEL data: high-latitude total intensity (open boxes:  Bloemen et
al.\ 1999, diamonds: Kappadath 1998, crosses:  Weidenspointner et
al.\ 1999).
\label{fig3} }
\end{figure}

\section{High latitude \grays and the size of the electron halo:} \label{grays}
We use our model for calculation of the Galactic contribution to the
high latitude diffuse \gray emission.  The high latitude \gray
intensity increases with halo size due to IC emission (though much less
than linearly due to electron energy losses), which allows us to put an
upper limit on the halo size.  Fig.~\ref{fig3} shows the \gray spectrum
towards the Galactic poles for $z_h = 4$ kpc, and 10 kpc.  $z_h = 10$
kpc is possible although the latitude profiles around 100 MeV
are then very broad and at the limit of consistency with EGRET
data.  Further the isotropic component would have to approach zero
above 300 MeV, so that this halo size can be considered an upper
limit.

If the halo size is 4--10  kpc as we argue, the contribution of
Galactic emission to the total at high latitudes is larger than
previously considered likely and has consequences for the derivation of
the diffuse extragalactic emission (e.g., Sreekumar et al.\  1998).  An
evaluation of the impact of our models on estimates of the
extragalactic spectrum is beyond the scope of the present work.

\section{Conclusions:}
The large electron/IC halo suggested here reproduces  well the latitude
variation of \gray emission from the plane to the poles, which can be
taken as support for the halo size deduced from
independent studies of cosmic-ray composition.  Halo sizes in the range
$z_h = 4-10$ kpc are favoured by both analyses.


\vspace{1ex}
\begin{center}
{\Large\bf References}
\end{center}
Berezinskii, V.S., et al.\ 1990, Astrophysics of Cosmic Rays
   (Amsterdam: North Holland) \\
Bloemen, H., et al.\ 1999, Astroph.\ Lett.\ Comm.\
   (3rd INTEGRAL Workshop), in press\\
Cline, D.B., \& Gao, Y.-T.\ 1992, \aap 256, 351\\
Dixon, D.D., et al.\ 1998, New Astronomy 3, 539\\
Kappadath, S.C. 1998, PhD Thesis, University of New Hampshire, USA\\
Kinzer, R.L., Purcell, W.R., \& Kurfess, J.D. 1999, \apj 515, 215\\
Lebrun, F., et al.\ 1999, Astroph.\ Lett.\ Comm.\
   (3rd INTEGRAL Workshop), in press\\
Moskalenko, I.V., \& Strong, A.W.  1999, submitted
  (astro--ph/9811296)\\ 
Page, D.N., \& Hawking, S.W.\ 1976, \apj 206, 1\\
Pohl, M., \& Esposito, J.A.  1998, \apj 507, 327\\
Ptuskin, V.S., \& Soutoul, A.  1998, \aap 337, 859\\
Skibo, J.G., et al.  1997, \apj 483, L95\\
Sreekumar, P., et al.\  1998, \apj 494, 523\\
Strong, A.W., \& Mattox, J.R.  1996, \aap 308, L21\\
Strong, A.W., \& Moskalenko, I.V.  1998, \apj 509, 212\\
Strong, A.W., \& Moskalenko, I.V.  1999, in ASP Conf.\ Ser.\ 171, 154\\
Strong, A.W., Moskalenko, I.V., \& Reimer, O.  1999, submitted
  (astro--ph/9811284)\\
Strong, A.W., et al.\  1998, Astroph.\ Lett.\ Comm.\
   (3rd INTEGRAL Workshop), in press\\
Valinia, A., \& Marshall, F.E.  1998, \apj 505, 134\\
Webber, W.R., \& Soutoul, A.  1998, \apj 506, 335\\
Weidenspointner, G., et al.\ 1999, Astroph.\ Lett.\ Comm.\
   (3rd INTEGRAL Workshop), in press\\

\end{document}